\documentclass[dvips,final]{acta}
\usepackage{supertabular,lscape,epsfig}
\usepackage{amssymb}
\usepackage{amsmath}

\SetPages{0}{0}
\SetVol{69}{2019}

\begin{document}

\begin{Titlepage}
\Title{On the relation between interstellar spectral features and reddening
}

\Author{Kre{\l}owski, J.}{Center for Astronomy, Nicholas Copernicus University, Gagarina 11, Pl-87-100 Toru{\'n}, Poland\\
        Rzesz{\'o}w University, al. T. Rejtana 16c, 35-959 Rzesz{\'o}w, Poland\\
        e-mail: jacek@astri.uni.torun.pl}

\Author{Galazutdinov, G.}{Instituto de Astronomia, Universidad Catolica del Norte, Av. Angamos 0610, Antofagasta, Chile\\
       Pulkovo Observatory, Pulkovskoe Shosse 65, Saint-Petersburg, Russia\\
       Special Astrophysical Observatory, Nizhnij Arkhyz, 369167, Russia \\
       e-mail: runizag@gmail.com}

\Author{Godunova, V.}{ICAMER Observatory, NAS of Ukraine, 27 Acad. Zabolotnoho Str.,
       03143,  Kyiv, Ukraine\\
       e-mail:v$\_$godunova@bigmir.net}

\Author{Bondar, A.}{ICAMER Observatory, NAS of Ukraine, 27 Acad. Zabolotnoho Str.,
       03143,  Kyiv, Ukraine\\
       e-mail:arctur.ab@gmail.com}

\end{Titlepage}

\Abstract{It is well known that interstellar spectral features correlate with
color excess E(B$-$V). This suggests that measuring intensities of these features allows one
to estimate reddening of stars. The aim of this paper is to check how tight intensities of interstellar diffuse bands (DIBs)
are related to the amount of extinction, measured using E(B$-$V).

We have measured equivalent widths of the strongest DIBs (centered at
$\lambda\lambda$ 5780.6, 5797.0, 6196.0, 6379.3, 6613.5, and 8620.7 \AA), as well
as of CH (near 4300.3 \AA) and CH$^+$ (near 4232.5 \AA) in high resolution,
high S/N ratio echelle spectra from several spectrographs.
The equivalent widths of the 8620 DIB in noisy spectra were measured using
a template, which was constructed using the high quality spectrum of BD+40\,4220.
DIB relations
with the color excess in the range 0.1--2.0 mag were examined.
Our careful analysis demonstrates that all the above mentioned interstellar
spectral features (except, perhaps, 6379 DIB) do correlate with E(B$-$V)
relatively tightly (with the Pearson's correlation coefficient of
0.8+). Moreover, the observed scatter is apparently not caused by
measurement errors but is of physical origin. We present several
examples where the strength ratios of a DIB/molecule to E(B$-$V)
are different than the average.
}{ISM: clouds -- ISM: dust, extinction: ISM: molecules -- lines and bands}

\section{Introduction}

Absorption features, originating in translucent interstellar
clouds, are being revealed due to the presence of the following
non-stellar phenomena in spectra of reddened OB stars:

\begin{itemize}
\item
Interstellar extinction, usually known as {\it reddening}, which
selectively attenuates starlight. It is commonly believed to be
caused by some solid particles of submicron size; however, since
the beginning of investigations of this phenomenon (Trumpler,
1930) some neutral (gray) extinction, caused by
relatively large dust particles, was proposed as well.
\item
Polarization -- also believed to be caused by dust particles when
the latter are non-spherical and oriented by e.g. magnetic field.
\item
Spectral lines of interstellar atomic gas, e.g., $H$ and $K$ doublet
of Ca{\sc ii}, discovered by Hartmann (1904). The
survey of UV interstellar lines (Field, 1974) proved
that heavy elements in the ISM are strongly depleted in comparison
to their abundances in stellar atmospheres, likely because of dust
formation out of heavy elements and of chemical reactions leading
to the formation of different (possibly complex) molecules.
\item
Molecular bands of simple radicals (CH, OH, NH, CN, C$_2$,
C$_3$, OH$^+$, SH), some of them were discovered and identified long ago
(McKellar 1941). Since the 1970s, rotational
emission features revealed the presence in the star forming regions of many complex
(currently $\sim$200) molecules with a dipole momentum;
\item
Diffuse interstellar bands (DIBs): their identification is the longest standing
unsolved
problem of the spectroscopy in general. The first two such features
were discovered in 1921 by Heger (1922). The
application of solid state detectors to DIB observations led to
discoveries of new features. Currently, the list of known DIBs
exceeds 500 entries (Galazutdinov et al., 2000, Hobbs et al., 2008, Fan et al., 2019); a majority
of them are very shallow. Even more importantly, the fine structure
(reminiscent of the rotational structure of polyatomic molecules)
has been detected in most of narrow DIBs (Kerr et al. 1998). Nearly
all conceivable forms of matter -- from hydrogen
anion to dust grains -- have already been proposed as DIB carriers,
so far with no generally accepted success. It should be noted that
their variable strength ratios suggest a variety of carriers, and
thus strongly support their molecular origin (for review, see Kre{\l}owski (2018)).
Publications of  Campbell et al. (2015, 2016, 2016a) re--started the discussion on whether
C$_{60}^{+}$  molecule may carry strong near--infrared spectral features of interstellar origin
seen at 9633 and 9577~~\AA.  Galazutdinov et al. (2017a, 2017b) disputed the identifications,
pointing out the problem of the variable strength ratio of these two strong features. The
problem of variable strength ratio remains unsolved also after the recent publication  of Cordiner et al. (2019)
based on HST spectra free of telluric contamination but not covering the 9633 DIB. 
\end{itemize}

As evidenced by Fitzpatrick and Massa (2007), it seems
well-established that interstellar absorption spectra, in particular extinction
curves,  may differ
from cloud to cloud. Sneden, Woszczyk \& Kre{\l}owski (1991) proposed already
a division of interstellar clouds into
$\sigma$ and $\zeta$ type objects. The interstellar spectra of such
objects are different showing variable strength ratios of molecular/atomic/diffuse 
features. In general diffuse bands intercorrelations are tight. 
These were recently
demonstrated by Bailey et al. (2016) 
who confirmed earlier results of Moutou et al.
(1999). Bailey et al. (2016) demonstrated also that in nearby, single cloud lines-of-sight
often these correlations weaken that results from observing $\sigma$ and $\zeta$ type
objects alone. Tight correlations follow usually adding up several clouds.
However, having spectra of high
enough resolution and S/N ratio, one can resolve individual Doppler
components in interstellar atomic and/or molecular lines. In cases
of diffuse bands, it is very difficult and possible only in very
specific occasions. When dealing with extinction and polarization
it is absolutely impossible. Since the extinction and polarization
curves may change when sightlines intersect several clouds
(in particular, the light beam may get depolarised by a subsequent
cloud), the only possibility to investigate these phenomena in
relation to other ones, is to observe stars when interstellar
atomic/molecular lines do not show any Doppler splitting. Such
objects are relatively scarce and may be found only by means of
high resolution spectral observations. High resolution echelle spectra can not only
allow one to distinguish
between single and multiple cloud cases. As already suggested
(Sneden, Woszczyk \& Kre{\l}owski, 1991), also the
linear polarization of interstellar spectra may vary together with the shape of the
extinction curve.

Amount and characteristics of dust vary greatly
from one interstellar cloud to another that poses difficulties in constructing
the distance scale in the Galaxy (e.g. Fitzpatrick \& Massa (2007), Siebenmorgen et al. (2014), etc).
In fact, distant objects may be reddened either by one or more translucent
interstellar clouds situated along the lines of sight towards them. Knowledge
of distances to cosmic objects is of basic importance for investigating their
physical properties, including absolute stellar magnitude. However, the latter
can be reciprocally used for distance estimation if being properly calibrated
to the spectrum and luminosity class of an observed object.

The most popular measure of extinction (reddening) is the color excess
E(B$-$V) = A$_B$$-$A$_V$ = (B$-$V)$_{obs}$ $-$ (B$-$V)$_o$,
where (B$-$V)$_{obs}$  is an object's observed color index, and (B$-$V)$_o$ is its intrinsic
value, whereas A$_B$ and A$_V$ are the total extinctions in the photometric $B$ (4400~\AA)
and $V$ (5500~\AA) bands, respectively. Another measure of extinction is the absolute
extinction A = A$_\lambda$/$A_V$, where A$_\lambda$ is the total extinction at wavelength
$\lambda$. As a rule, color excess grows with increasing absolute extinction.

Extinction effects (e.g. reddening) are most often observed in spectra of early-type
stars due to their high intrinsic brightness, which allows one to observe distant,
heavily reddened objects. Extinction has for a very long time been considered to be
correlated with spectral features of interstellar origin, in particular with DIBs, which are best seen especially towards the above
mentioned early-type objects (of spectral types O or B) strongly radiating
in the optical region with relatively few purely stellar absorption features.

Years ago, Merrill \& Wilson (1938) established the presence of the correlation between the
6284 DIB and color excess. Later Greenstein \& Aller(1950)
demonstrated a similar effect for the strongest DIB centered at $\lambda$ 4430 \AA.
An extensive examination of eight DIBs and their correlations was performed by
Friedman et al.(2011). Furthermore, behavior of interstellar features in spectra of
reddened stars was a subject of intent scientific research by
Friedman et al.(2011), Kashuba et al.(2016), York et al.(2014). New, infrared DIBs have recently been
described by Hamano et al.(2016), Galazutdinov et al.(2017a). For a recent review see Kre{\l}owski (2018).
Results of these and other studies may suggest application of interstellar
features, such as diffuse interstellar bands, to estimate color excesses and,
consequently, total extinction.

\section{Investigations of the 8620 DIB as a spectroscopic tracer of extinction}

The diffuse interstellar band at $\lambda$ 8620.7 \AA\ (the so-called ``Gaia DIB'')
belongs evidently to the class of reasonably broad features. It is the only unambiguous DIB in the wavelength range
8450 -- 8720~\AA\ of the Radial Velocity Spectrometer (RVS) aboard Gaia (www.cosmos.esa.int/web/gaia).

For another similar one
(at $\lambda$ 5780.6 \AA; a bit narrower than the 8620 one) Kre{\l}owski \& Westerlund(1988) proved that the DIB
intensity can change by a factor of 3 with the same E(B$-$V) while comparing spectra
of two bright stars: $\sigma$~Sco and $\zeta$~Oph (see Fig. 1 of Kre{\l}owski \& Westerlund(1988)). This led to the division of
translucent interstellar clouds into $\sigma$ and $\zeta$ cases being now in common use.
A considerable growth in the number of studies of the 8620 DIB and its correlation
with reddening has been specifically driven by the Gaia space mission.
We provide here a short overview of some recently published works.
Wallerstein et al.(2007) measured the equivalent width (EW) of this DIB in spectra of 64 stars (recorded with different resolutions) with
color excesses of up to 3.17 mag and found a relatively good correlation between
the 8620 DIB intensity and E(B$-$V). However, the lack of the EW errors reduces the significance
of the reported conclusions. The method of equivalent width measurements does not
bear scrutiny either: ``equivalent widths of the DIBs and K{\sc i} were measured via
Gaussian fits to the observed line profiles'' (page 1272). It is important to
underscore that the Gaussian fit is not well applicable for irregular profiles of
diffuse bands. Moreover, Wallerstein et al. used in many cases compilations of
measurements, done by somebody else. This creates additional uncertainties as different researchers set
continua over DIBs in different fashions. Such averages suffer an excessive scatter which reduces 
the significance of the results.

Munari et al.(2008) proposed to use the 8620 DIB as a tool to measure the amount of reddening.
The equivalent width of this DIB seems (according to Munari et al.) to correlate very tightly with E(B$-$V) as
seen in Fig. 2 of
Munari et al.(2008). However, it is of interest to consider how certain is the E(B$-$V)
estimate based on the EW(8620) intensity? Indeed, Munari's relation is based on the
76 low resolution (R=7500) and relatively low S/N ratio (as low as 50 only) spectra of
68 stars. This may smear out some local differences that makes the relation
uncertain.

Kos et al.(2013) confirmed the general correlation between the equivalent width of
the 8620 DIB  and the reddening  in a rather statistical way where both the
equivalent width and the reddening magnitude were estimated for a group of targets
in a certain volume instead of individual estimations. In the next paper Kos et al.(2014)
 combined information from nearly 500,000 stellar spectra obtained by the massive
spectroscopic survey RAVE to produce a pseudo-3D map of the strength of the 8620 DIB
covering the nearest 3 kpc from the Sun, and showed that the 8620 DIB  carrier has
a significantly larger vertical scale height than the dust. It should be noted that
the method applied and the quality of the analyzed data (R=7500 with low S/N ratio)
can provide very general conclusions only.

Ma\'{\i}z Apell\'{a}niz et al.(2015) studied three objects in the Berkeley 90 cluster.
The authors applied the Gauss fit method to measure the equivalent width of
the 8620 DIB. As a result, the following conclusion was drawn: ``DIB  $\lambda$ 8621.20 \AA,
present in the Gaia band, is a good example of a strong DIB that is expected to
correlate poorly with extinction because it appears to be highly depleted in the
dense ISM''. However, the low resolution as well as the quality of their spectra
make  the conclusions uncertain. In particular, low resolution does not allow
to distinguish with certainty different Doppler components.
Damineli et al.(2016) measured the equivalent width of the 8620 DIB using low
resolution spectra (R $\le$ 15000) obtained for 11 bright members from the Westerlund 1
stellar cluster and for 12 objects along other Galactic directions. The authors
derived a relation EW(8620) vs. E(B$-$V), which extends Munari et al.(2008)'s relation
to the non-linear regime (A$_V$ $\ge$ 4 mag).

\begin{figure}
\includegraphics[width=12cm]{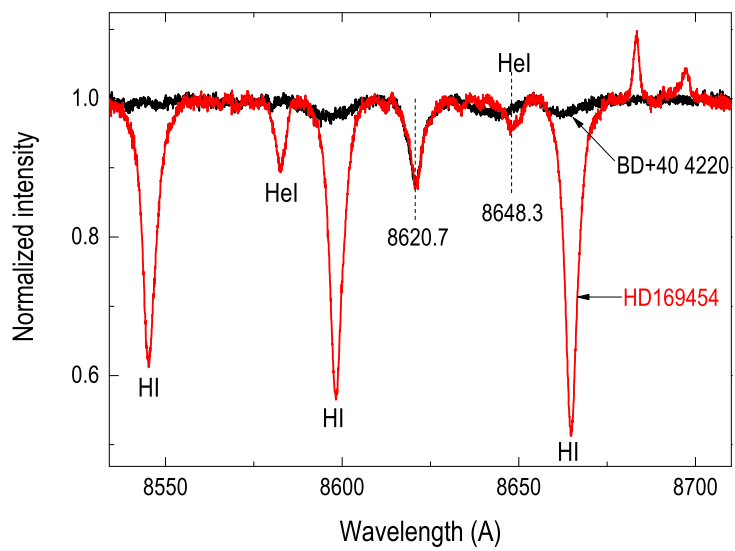}
\caption{DIB 8620 and stellar lines in the spectra of two heavily
reddened stars normalized to the depth of the 8620 DIB. Note the Paschen and
He{\sc i} lines. The vertical dashed lines indicate the interstellar
features, mentioned by Jenniskens \& Desert (1994).}
\end{figure}

The Gaia DIB is situated in the wavelength range populated with
the hydrogen Paschen lines. In B type spectra one can also observe lines
of helium and some other elements, as shown in Fig. 1. This plot
demonstrates also that the 8648 DIB, mentioned by Jenniskens \& Desert (1994), is most probably
a stellar helium line. Munari et al.(2008) mention two He{\sc i} lines
at 8648.258~\AA\ and 8650.811~\AA\ in the vicinity of the postulated 8648 DIB.
Besides, the feature (DIB 8648) does not correlate with reddening.
The line is blue-shifted (in  the spectrum of BD+40 4220) in relation to the 8620 DIB, which is
evident and strong. BD+40 4220 is a
heavily reddened, rapidly rotating, very hot star and, therefore, it is very
useful to separate stellar and interstellar features. DIB 8620 is relatively broad feature: indeed, its typical full
width at the half maximum (FWHM) is about 130 km/s. For example, FWHM of
the major DIB 5780 is $\sim$ 110 depending on the line of sight. Hereby, Fig. 1
proves that the DIB 8620  is the only unambiguous broad and strong spectral feature of
interstellar origin in the depicted spectral range.

Does the 8620 DIB intensity provide a better estimate of E(B$-$V)
than other DIBs? The first problem is the spectral range of
near-IR. In this range the quantum efficiency of CCD matrices is
much lower than in the visual one that makes the S/N
ratio lower. Moreover, one can expect some stellar contaminations,
especially in the continuum around the 8620 DIB, that is also
demonstrated in Fig.~1 of Munari et al.(2008).
Owing to the complexity of the 8620 DIB profile, we decided to measure
it by an original method developed in DECH package$\footnote{$http://www.gazinur.com/DECH-software.html$}$: we have used 8620 DIB observed in the spectrum of
the heavily reddened and rapidly rotating object BD+40~4220 as a template.
To measure this DIB in other targets we have rescaled manually the depth of the template
to match the feature in other targets. BD+40~4220 was observed using the
ESPaDOnS spectrograph attached at the 3.6m CFHT. The spectrograph
is especially well designed for observations in near-IR range. Since the DIB
is broad -- its profile is not affected by the Doppler splitting in any object.

Fig. 2
demonstrates how the measurements have been done. The
high S/N profile of 8620 DIB  (seen in BD+40 4220) was rescaled (made shallower or deeper)
to match the same feature observed in other spectra. Fig. 1 shows how it works. Then, the equivalent width of the rescaled template was
measured and attributed to the studied target. This allowed us to avoid the influence of stellar contaminations and reduce the noise effects. The
8620 DIB is a broad feature, and its profile shape does not depend on the spectral resolution varying in our sample. The DIB widths and profiles are not exactly 
constant but they usually do not differ much in cases of broad features, like 8620.

\begin{figure}
\includegraphics[width=12cm]{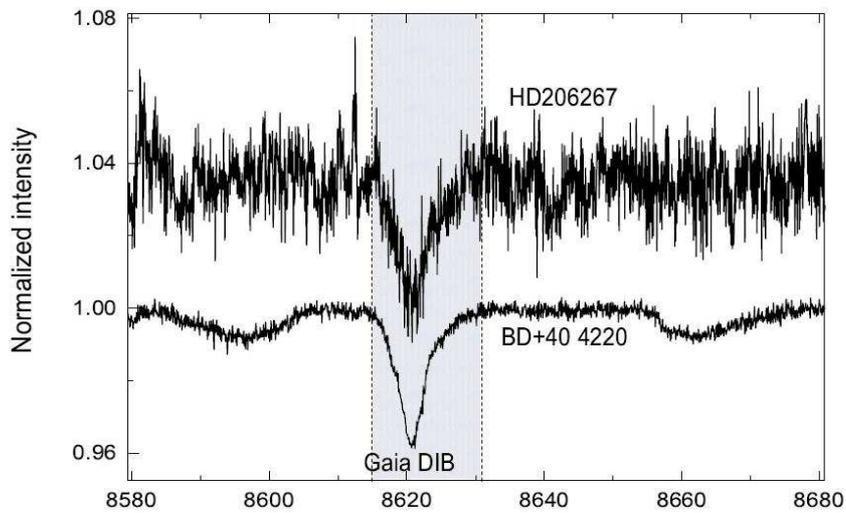}
\caption{
An illustration of the method we have applied for measurements of the equivalent width of  8620 DIB in noisy spectra.
Profile of 8620 DIB with a high S/N ratio (seen in BD+40 4220) was rescaled (made more shallow than in the observed spectrum)
to match the depth of the same interstellar feature observed in the spectrum of HD206267.
Then, the equivalent width of the rescaled template was measured and attributed to the studied target. In the figure, spectra are vertically displaced for clarity.
}
\label{measure}
\end{figure}

\begin{figure}
\includegraphics[width=12cm]{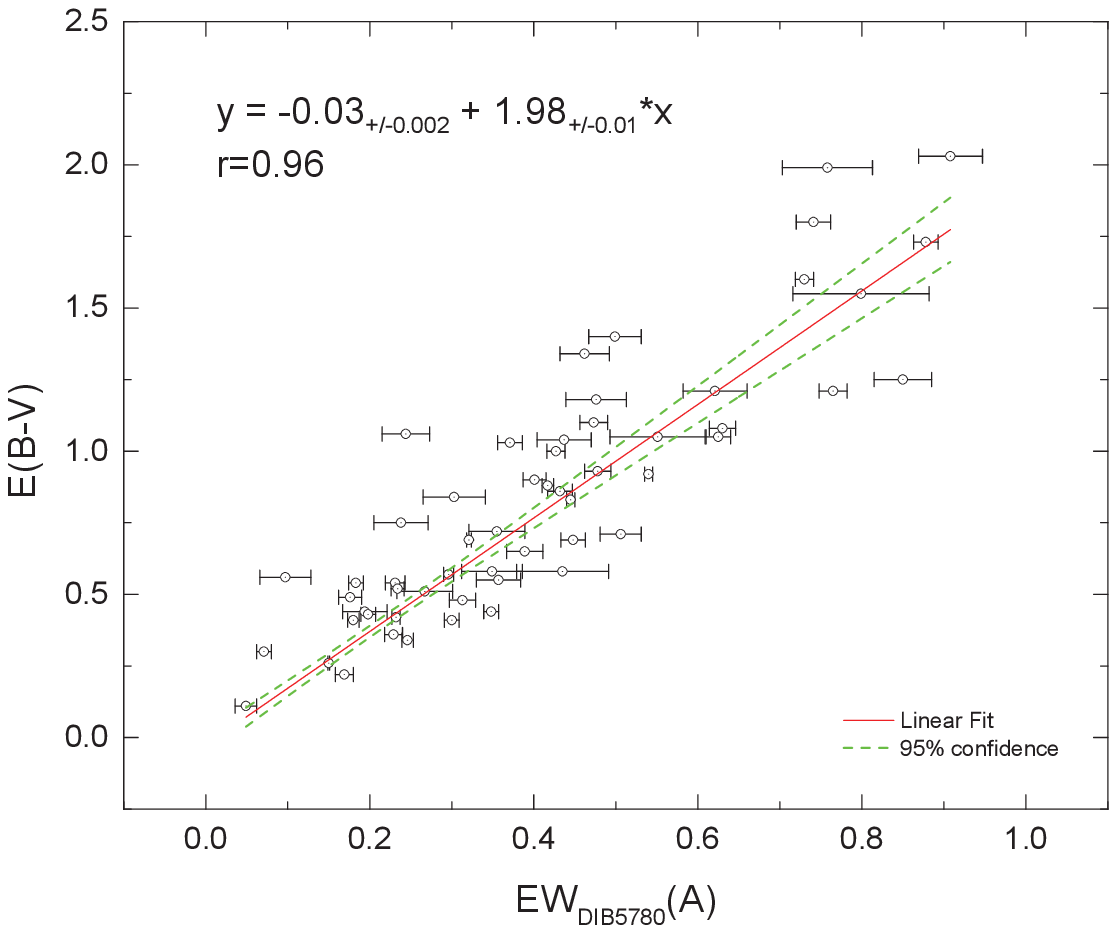}
\bigskip
\caption{EW(5780) as the proxy of E(B-V). Also in this
case (like in Fig. 4) the observed scatter is most likely grounded in physical
parameters of the intervening clouds.}
\label{corel5780}
\end{figure}

\section{Observations}

Our analyzed data were acquired using the following four \'{e}chelle spectrographs:
\begin{itemize}
\item{
BOES (Bohyunsan Echelle Spectrograph) of the Korean National Observatory
(Kim et al. (2007)) is installed at the 1.8m telescope of the Bohyunsan
Observatory in Korea. The spectrograph has three observational modes
allowing resolving powers of 30,000, 45,000 and 90,000. The lowest
resolution, which enables observations of rather faint, heavily
reddened, distant objects, was used in most of cases. In any mode, the
spectrograph covers the whole spectral range from $\sim$3500 to
$\sim$10,000 \AA, divided into 75 -- 76 spectral orders;}
\item{
MAESTRO (MAtrix Echelle SpecTROgraph) is attached to the 2~m
telescope at the Terskol Observatory (the North Caucasus). It is a
three branch cross-dispersed {\'e}chelle spectrograph installed at the Coud{\'e}
focus (F/36) of the telescope. It was designed for stellar spectroscopy
using high resolutions from R = 45,000 to 190,000 in the spectral range
3500 -- 10000 \AA. The lowest resolution mode (sufficient for our
programme) allows one to reach spectra of objects as faint as $\sim$10$^m$
with a sufficient ($\sim$100) signal-to-noise ratio;}
\item{
CAFE spectrograph (Calar Alto high-Resolution facility, Southern Spain)
fed by 2.2m telescope provides high resolution {\'e}chelle spectra over 3700--9100~\AA. The resolution
of the spectrograph is up to R = 62,000; the whole range is divided into 92 orders;}
\item{
ESPaDOnS spectrograph (Echelle SpectroPolarimetric Device
for the Observation of Stars) is the bench-mounted high-resolution {\'e}chelle
spectrograph/spec\-tro\-pola\-ri\-meter attached to the 3.58~m Canada-France-Hawaii
telescope at Mauna Kea (Hawaii, US). It is designed to obtain a complete optical
spectrum in the range from 3,700
to 10,050~~\AA. The whole spectrum is divided into 40 {\'e}chelle orders. The
resolving power is about 68,000. The ESPaDOnS is especially useful for observations
in near-IR since the high altitude of the
observatory allows one to minimize the telluric contaminations.}
\end{itemize}

The spectra from BOES were gathered during the period from 2004 to 2015,
with PI G.~Galazutdinov. The spectra from ESPaDOnS were obtained during the runs
05Ao5 (in 2010, with PI B.~Foing) and 15AD83 (in 2015, with PI G.~Walker).
Spectra from MAESTRO were acquired between 1996 and 2016 by the authors. Spectra from CAFE
were collected in February 2007 by Y. Beletsky.

\begin{figure}
\includegraphics[width=12cm]{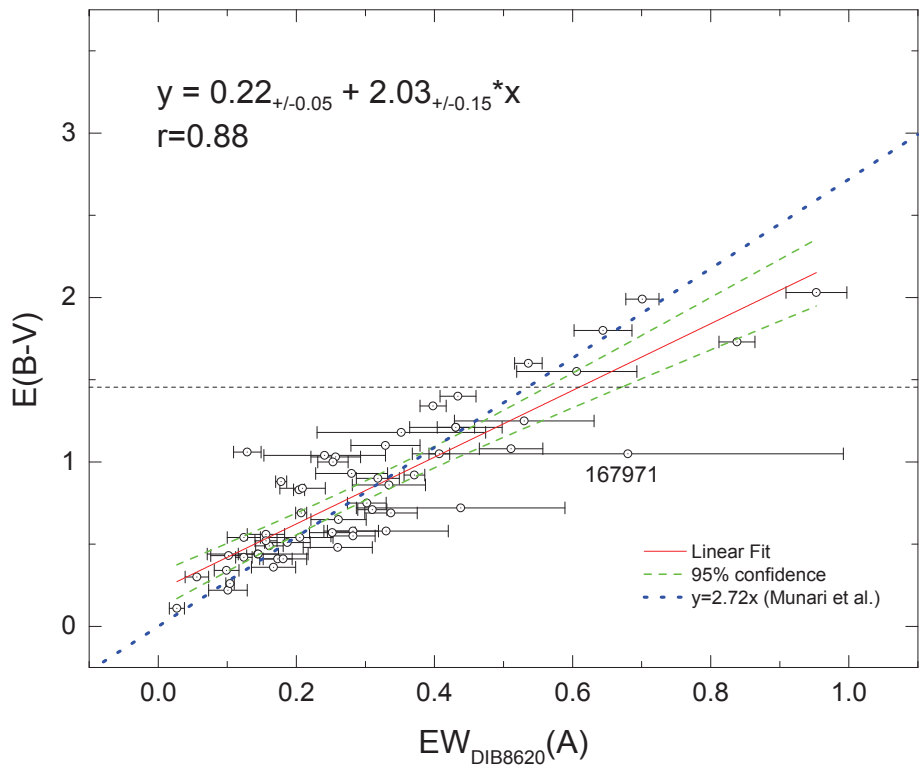}
\caption{E(B$-$V) plotted as a function of EW(8620) for our sample of
spectra. Horizontal dashed line sets the E(B$-$V) limit for the sample of
Munari et al. (2008); his fit to the depicted points is shown together
with ours. (Note: samples are not identical.)
}
\label{corel8620}
\end{figure}

The above mentioned spectra are of various quality, therefore,
those which do not allow reliable measurements were not involved in
our study. Our sample includes those spectra which mostly have high
enough S/N ($\ge$150) and where the feature at $\lambda$ 8620.7 \AA\ can
be reliably traced. All spectra have been reduced by the authors.

\begin{table*}
\caption{Equivalent widths of DIBs and molecules (in m\AA).}
\label{starsSK}{\footnotesize
\begin{tabular}{lrrrrrrrrrr@{}}
\hline
Star& E$_{B-V}$& 8620& 5780& 5797& 6196& 6379& 6614& CH,& CH$^+$, \\
& & & & & & & &(4300 \AA) &(4232 \AA) &  \\
\hline \\
BD+59451$^b$&   0.86&  334$\pm$53&   432$\pm$15&  134$\pm$ 5&   48$\pm$ 2&  88$\pm$ 3&  182$\pm$ 5& 40.0$\pm$ 2.0&  18.0$\pm$ 2.0 \\
BD+59456$^b$&   0.69&  337$\pm$38&   448$\pm$15&  121$\pm$ 6&   46$\pm$ 2&  89$\pm$ 3&  202$\pm$ 8& 40.0$\pm$ 3.0&  23.0$\pm$ 2.0 \\
BD+60594$^b$&   0.58&  282$\pm$37&   349$\pm$37&  138$\pm$13&   40$\pm$ 2&  69$\pm$ 2&  162$\pm$ 5& 25.0$\pm$ 2.0&  17.0$\pm$ 1.0 \\
BD+404220$^e$&  1.99&  701$\pm$24&   758$\pm$55&  239$\pm$25&   96$\pm$10&  95$\pm$10&  306$\pm$19& 78.0$\pm$ 6.0&  86.0$\pm$ 8.0 \\
BD+582580$^b$&  1.08&  511$\pm$46&   630$\pm$16&  212$\pm$ 9&   82$\pm$ 2&  89$\pm$ 2&  291$\pm$11& 35.0$\pm$ 5.0&  21.0$\pm$ 3.0 \\
BD+592735$^b$&  1.25&  530$\pm$101&  850$\pm$35&  282$\pm$15&  117$\pm$10& 189$\pm$10&  424$\pm$25& 60.4$\pm$19.8&  45.7$\pm$15.4 \\
CygOB2\_7$^b$&  1.80&  644$\pm$42&   741$\pm$21&  197$\pm$10&   89$\pm$10&  95$\pm$ 9&  327$\pm$20& 65.5$\pm$11.9&  93.7$\pm$32.6 \\
CygOB2\_8A$^b$& 1.60&  536$\pm$20&   730$\pm$11&  175$\pm$ 5&   93$\pm$ 8&  88$\pm$ 5&  336$\pm$16& 43.0$\pm$ 3.0&  83.0$\pm$ 3.0 \\
CygOB2\_11$^b$& 1.73&  838$\pm$26&   878$\pm$15&  179$\pm$10&  112$\pm$16&  87$\pm$ 8&  366$\pm$22& 43.1$\pm$12.1&  94.5$\pm$20.9 \\
CygOB2\_12$^e$& 2.03&  953$\pm$44&   908$\pm$39&  330$\pm$37&  115$\pm$19& 140$\pm$19&  391$\pm$37&     $-  $    &      $-  $     \\
13256$^b$&      1.21&  431$\pm$67&   621$\pm$39&  223$\pm$18&   90$\pm$ 6&  93$\pm$ 6&  297$\pm$21& 46.0$\pm$ 9.0&      $-  $     \\
14956$^b$&      0.88&  178$\pm$8&    417$\pm$ 7&  145$\pm$ 3&   45$\pm$ 1& 105$\pm$ 1&  219$\pm$ 4& 37.2$\pm$ 2.2&  26.7$\pm$ 3.2 \\
15497$^b$&      0.83&  204$\pm$8&    445$\pm$ 5&  142$\pm$ 2&   43$\pm$ 1&  65$\pm$ 2&  232$\pm$ 4& 44.5$\pm$15.9&  60.8$\pm$19.4 \\
15785$^b$&      0.65&  261$\pm$40&   389$\pm$22&  150$\pm$11&   44$\pm$ 5&  96$\pm$ 5&  220$\pm$10& 26.2$\pm$ 1.8&  23.4$\pm$ 4.0 \\
21291$^t$&      0.44&  144$\pm$73&   194$\pm$27&   67$\pm$13&   26$\pm$ 6&  34$\pm$ 8&   95$\pm$18& 16.1$\pm$ 4.1&   7.2$\pm$ 2.6 \\
24534$^b$&      0.56&  156$\pm$27&    97$\pm$31&   73$\pm$ 5&   17$\pm$ 5&  56$\pm$ 8&   75$\pm$10& 23.7$\pm$ 1.9&   3.7$\pm$ 1.8 \\
31327$^c$&      0.54&  124$\pm$24&   183$\pm$ 9&   77$\pm$ 4&   30$\pm$ 2&  70$\pm$ 4&  111$\pm$ 6& 25.7$\pm$ 5.0&  23.3$\pm$ 6.2 \\
32991$^c$&      0.41&  173$\pm$21&   180$\pm$ 7&   68$\pm$ 3&   20$\pm$ 3&  43$\pm$ 2&   68$\pm$ 4& 20.3$\pm$ 2.3&  15.4$\pm$ 1.8 \\
34078$^e$&      0.49&  161$\pm$20&   176$\pm$14&   57$\pm$ 6&   20$\pm$ 4&  17$\pm$ 4&   70$\pm$ 9& 54.8$\pm$ 3.9&  37.4$\pm$ 4.2 \\
36371$^c$&      0.41&  181$\pm$34&   300$\pm$ 9&   90$\pm$ 3&   38$\pm$ 2&  81$\pm$ 4&  141$\pm$ 4& 16.5$\pm$ 3.0&  15.2$\pm$ 3.5 \\
36861$^b$&      0.11&   27$\pm$11&    49$\pm$13&   20$\pm$ 4&    5$\pm$ 2&   7$\pm$ 2&   16$\pm$ 6&  1.8$\pm$ 0.2&   0.9$\pm$ 0.1 \\
41117$^c$&      0.44&  145$\pm$29&   348$\pm$ 9&  117$\pm$ 4&   34$\pm$ 1&  64$\pm$ 2&  151$\pm$ 4& 14.0$\pm$ 2.2&  19.9$\pm$ 6.8 \\
42087$^c$&      0.34&   99$\pm$18&   246$\pm$ 7&   95$\pm$ 4&   30$\pm$ 2&  75$\pm$ 6&  105$\pm$ 4& 14.3$\pm$ 3.9&   7.5$\pm$ 2.8 \\
43384$^b$&      0.58&  330$\pm$90&   435$\pm$56&  130$\pm$29&   47$\pm$12&  88$\pm$15&  199$\pm$24& 16.1$\pm$ 6.4&  32.0$\pm$ 8.5 \\
46202$^b$&      0.48&  260$\pm$50&   313$\pm$16&   88$\pm$ 8&   36$\pm$ 3&  52$\pm$ 3&  153$\pm$11& 15.9$\pm$ 3.4&  15.7$\pm$ 4.8 \\
145502$^e$&     0.22&  101$\pm$28&   169$\pm$11&   34$\pm$ 5&   16$\pm$ 3&  28$\pm$ 3&   60$\pm$ 8&  3.6$\pm$ 0.5&   6.6$\pm$ 1.1 \\
147165$^b$&     0.36&  167$\pm$32&   229$\pm$11&   37$\pm$ 4&   18$\pm$ 1&  29$\pm$ 2&   65$\pm$ 5&  3.1$\pm$ 0.1&   5.3$\pm$ 0.3 \\
147889$^b$&     1.03&  257$\pm$36&   371$\pm$15&  156$\pm$ 8&   34$\pm$ 3&  81$\pm$ 4&  173$\pm$ 7& 49.8$\pm$ 1.1&  27.7$\pm$ 1.2 \\
147933$^e$&     0.43&  102$\pm$26&   198$\pm$ 9&   51$\pm$ 5&   16$\pm$ 2&  28$\pm$ 3&   66$\pm$ 6& 17.9$\pm$ 0.4&  13.1$\pm$ 1.4 \\
149757$^e$&     0.30&   56$\pm$17&    71$\pm$ 9&   33$\pm$ 3&   11$\pm$ 2&  18$\pm$ 2&   44$\pm$ 4& 17.9$\pm$ 0.1&  23.5$\pm$ 2.1 \\
167971$^t$&     1.05&  680$\pm$312&  551$\pm$58&  163$\pm$25&   78$\pm$24& 109$\pm$ 8&  254$\pm$28& 29.0$\pm$ 2.0&  51.7$\pm$ 2.8 \\
168607$^b$&     1.55&  606$\pm$87&   799$\pm$83&  261$\pm$35&  100$\pm$ 4& 159$\pm$ 2&  393$\pm$13& 47.1$\pm$ 2.9&  73.0$\pm$ 1.8 \\
169454$^e$&     1.10&  329$\pm$50&   473$\pm$17&  159$\pm$ 8&   61$\pm$ 6& 102$\pm$ 6&  216$\pm$17& 29.8$\pm$ 0.5&  18.2$\pm$ 0.4 \\
173438$^b$&     1.00&  253$\pm$22&   427$\pm$11&  209$\pm$ 8&   42$\pm$ 2& 100$\pm$ 3&  229$\pm$10& 34.0$\pm$ 3.0&  20.0$\pm$ 2.0 \\
183143$^e$&     1.26&  431$\pm$27&   765$\pm$17&  195$\pm$ 8&   93$\pm$ 4& 111$\pm$ 5&  361$\pm$16& 38.7$\pm$ 3.0&  49.0$\pm$ 2.8 \\
185859$^b$&     0.57&  252$\pm$33&   296$\pm$ 6&  148$\pm$ 3&   42$\pm$ 1& 142$\pm$ 5&  213$\pm$ 3& 21.0$\pm$ 1.0&  18.0$\pm$ 2.0 \\
186745$^b$&     0.93&  280$\pm$52&   478$\pm$16&  198$\pm$ 5&   60$\pm$ 2& 139$\pm$ 2&  278$\pm$ 7& 40.0$\pm$ 2.0&  46.7$\pm$ 5.5 \\
190603$^t$&     0.72&  438$\pm$151&  355$\pm$34&   89$\pm$17&   36$\pm$ 8&  82$\pm$22&  109$\pm$27& 15.2$\pm$ 1.6&  31.4$\pm$ 2.7 \\
190918$^b$&     0.42&  124$\pm$12&   232$\pm$ 5&   58$\pm$ 4&   27$\pm$ 2&  14$\pm$ 1&   92$\pm$ 4&  9.9$\pm$ 5.8&  14.3$\pm$ 3.1 \\
192163$^b$&     0.26&  104$\pm$6&    150$\pm$ 1&   52$\pm$ 2&   26$\pm$ 1&  14$\pm$ 1&   76$\pm$ 2&  7.5$\pm$ 1.4&  17.1$\pm$ 2.1 \\
193793$^b$&     0.69&  207$\pm$8&    321$\pm$ 3&   94$\pm$ 1&   41$\pm$ 1&  32$\pm$ 1&  161$\pm$ 4& 22.4$\pm$ 2.0&  53.4$\pm$ 8.1 \\
194279$^t$&     1.18&  352$\pm$122&  476$\pm$37&  127$\pm$12&   58$\pm$15& 105$\pm$10&  201$\pm$34& 33.4$\pm$ 9.9&  47.0$\pm$15.9 \\
204827$^b$&     1.06&  129$\pm$20&   244$\pm$29&  187$\pm$13&   37$\pm$ 2&  91$\pm$ 1&  163$\pm$ 4& 68.7$\pm$ 3.7&  31.9$\pm$ 3.7 \\
206267$^t$&     0.54&  205$\pm$48&   231$\pm$12&   85$\pm$ 5&   26$\pm$ 1&  40$\pm$ 3&  122$\pm$ 6& 21.0$\pm$ 1.0&  14.0$\pm$ 1.0 \\
207538$^b$&     0.51&  187$\pm$33&   267$\pm$34&  166$\pm$17&   33$\pm$ 1&  96$\pm$ 2&  168$\pm$ 6& 30.4$\pm$ 1.8&   7.3$\pm$ 1.3 \\
208501$^b$&     0.75&  302$\pm$28&   238$\pm$33&  112$\pm$20&   31$\pm$ 2&  63$\pm$ 3&  113$\pm$ 4& 37.0$\pm$ 4.2&  12.6$\pm$ 5.2 \\
210839$^t$&     0.53&  156$\pm$16&   234$\pm$ 8&   78$\pm$ 3&   31$\pm$ 1&  53$\pm$ 2&  140$\pm$ 4& 21.4$\pm$ 1.9&   9.3$\pm$ 0.4 \\
217035$^b$&     0.71&  310$\pm$21&   506$\pm$25&  142$\pm$10&   50$\pm$ 2&  88$\pm$ 2&  211$\pm$ 6& 27.0$\pm$ 2.0&  25.0$\pm$ 1.0 \\
217086$^t$&     0.92&  371$\pm$15&   540$\pm$ 5&  156$\pm$ 2&   59$\pm$ 2&  87$\pm$ 2&  266$\pm$ 4& 39.0$\pm$ 3.0&  44.1$\pm$ 4.0 \\
219287$^b$&     1.05&  407$\pm$15&   625$\pm$15&  166$\pm$ 7&   81$\pm$ 2&  73$\pm$ 2&  261$\pm$ 7& 33.0$\pm$ 3.0&  21.5$\pm$ 2.0 \\
226868$^b$&     0.90&  318$\pm$31&   401$\pm$14&  125$\pm$ 5&   45$\pm$ 2&  68$\pm$ 2&  171$\pm$ 4& 41.0$\pm$ 2.0&  51.0$\pm$ 2.0 \\
228712$^b$&     1.34&  398$\pm$19&   462$\pm$30&  123$\pm$12&   57$\pm$ 7&  57$\pm$ 5&  189$\pm$15& 37.0$\pm$ 1.0&  34.0$\pm$ 2.0 \\
228779$^b$&     1.40&  434$\pm$26&   499$\pm$32&  194$\pm$14&   62$\pm$ 1& 106$\pm$ 2&  268$\pm$ 6& 54.4$\pm$ 6.6&  51.1$\pm$ 13.6 \\
235825$^b$&     0.55&  282$\pm$32&   357$\pm$27&   98$\pm$15&   28$\pm$ 2&  52$\pm$ 3&  134$\pm$ 4&  9.8$\pm$ 4.9&   7.0$\pm$ 2.4 \\
254577$^b$&     1.04&  241$\pm$88&   437$\pm$33&  145$\pm$14&   52$\pm$ 6&  84$\pm$ 5&  208$\pm$13& 38.6$\pm$ 6.2&  30.3$\pm$ 8.0 \\
281159$^b$&     0.84&  209$\pm$33&   303$\pm$38&   99$\pm$15&   31$\pm$ 2&  56$\pm$ 2&  152$\pm$ 6& 40.0$\pm$ 2.0&  38.0$\pm$ 1.0 \\
\hline
\end{tabular}}
\end{table*}

\section{Results and discussion}

We have measured the equivalent widths of the most prominent DIBs at $\lambda\lambda$
 8620.7, 5780.6, 5797.0, 6196.0, 6379.3, and 6613.5 \AA, as well as of the two strongest
molecular features, namely CH (near 4300.3 \AA) and CH$^+$ (near
4232.5 \AA), in the spectra of 56 reddened stars, which are distributed around
the Milky Way disc. As mentioned, our
sample includes only spectra with S/N $\ge$150.

All the measurements of equivalent width and color excess
E(B$-$V) are collected in Table~1. Here, letters `b', 'c', `e', and
 `t' stand for BOES, CAFE, ESPaDOnS, and MAESTRO spectrographs, respectively.
(B$-$V)'s of selected targets were taken from the SIMBAD database;
their intrinsic values allowed us to calculate color excesses E(B$-$V)
(see Papaj et al.(1993)).

Comparison of our values of E(B$-$V) and EW, calculated for DIBs at
$\lambda$ 5780.6, 5797.0, 6196.0, and 6613.5 \AA, with those
presented in Friedman et al.(2011) indicates that both sets sufficiently
agree. The relation between EW(5780) and E(B$-$V) given by
Friedman et al.(2011) (the correlation coefficient is r=0.82) agrees with our 
one (r=0.96) (Fig.~3) proxy
 inside the calculated errors. The
differences of 0.01 -- 0.05 in E(B$-$V) are probably due to spectral
stellar classification errors. A nearly systematic mismatch for EWs
at $\lambda$ 5797.0 \AA\ (namely, Friedman's values are larger than ours)
is caused by inclusion in measurements of the blue wing or rather
the additional DIB at $\lambda$ 5795 \AA, the latter being
correlated rather with the neighbouring 5780 DIB than with 5797 DIB. This
is a good example of the risk of using measurements from different
sources for a compilation.

We used the results obtained (and presented here for the first time) to check the relation between EW(8620) and
E(B$-$V) reported by Munari et al.(2008). Our project is the first one  based
on high resolution spectra. Fig.~4 depicts E(B$-$V) plotted
as a function of EW(8620) for our sample of spectra.
Our relation is reasonably similar to that of Munari et al.(2008) but some differences
bring one to a number of questions: Is the observed scatter a result of
measurement errors? Or the fact that we used another sample of objects plays an
important role? And is the 8620 DIB a unique feature,
especially tightly related to E(B$-$V)?

\begin{figure}
\includegraphics[width=12cm]{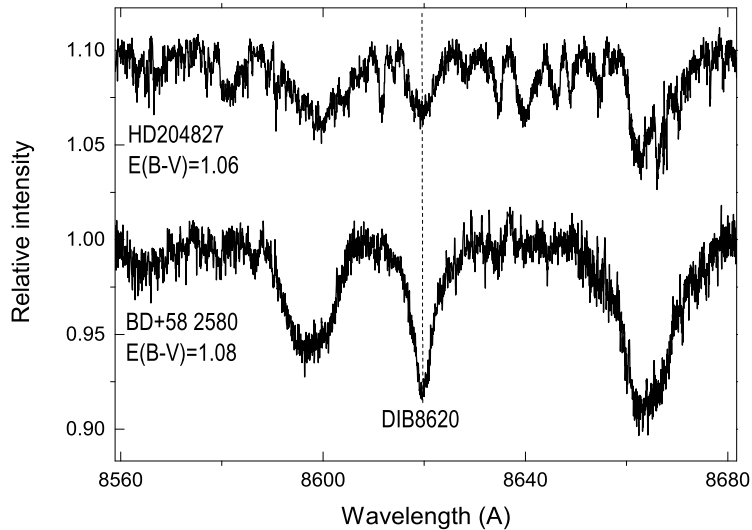}
\caption{Spectra (parts) of two identically reddened stars with the marked 8620 DIB.
The great apparent difference of the DIB's strengths is evident.}
\label{comp8620}
\end{figure}

Toward answering these questions, we shall expound below our findings,
based upon an analysis of both 8620 DIB and some other DIBs.

Fig.5 demonstrates the same parts of two spectra
acquired using the BOES spectrograph of the Korea National
Observatory. Both objects are almost identically reddened. However,
8620 DIB is $\sim$2.5 times stronger in BD+58~2580. It is a matter
of fact that the observed intensities of 8620 DIB can be
different in spectra of objects with the same E(B$-$V). This
resembles the behaviour of the (similarly broad) 5780\AA\ DIB reported by
Kre{\l}owski \& Westerlund (1988). Therefore, the scatter, which is seen in Fig.~4,
apparently follows physical differences between intervening clouds.
It thus suggests that using EW(8620) to estimate E(B$-$V) in any
individual case may be risky. In fact, a very similar relation
between equivalent width and colour excess may be found for another
broad DIB, namely 5780 DIB (Fig. 3).

A relation between EW(5780) and E(B$-$V) was recently published by
Raimond et al.(2012). Their relation, which suffers a very broad
scatter (their Fig.~3), is somewhat different than that of us. It
is very likely an effect of using nearby, mostly late B type
stars. Likely while considering relatively nearby objects the inhomogeneity 
of the interstellar medium plays an important role. In such cases an DIB 
intensity can hardly play the role of E(B$-$V) proxy.

Raimond et al.(2012) calculated the distances from the Hipparcos trigonometric
parallaxes, given in the SIMBAD database. These parallaxes are,
likely in many cases, incorrect due to large errors. For instance, the
authors give the distance to HD165052 as big as 6667 pc.  The Gaia DR1
parallax of HD165052 is 0.66$\pm$0.31 mas that leads to a distance of 1515 pc.
The new Gaia DR2 one is 0.78$\pm$0.05 that leads to the distance of 1280 pc.
It seems worth of being mentioned that our method, based on
interstellar Ca{\sc ii} line intensities (Megier et al.  2005), gives a
distance 1250 pc to the same star.

Limiting the range of EW(5780) to  a more narrow one led to the
 higher scatter. For example, in Fig. 3 of Friedman et al.(2011), where the scatter is  quite high,  the E(B$-$V) range is limited within 0 -- 0.7 mag only.
Our Fig.3 shows a much lower scatter while the E(B$-$V) is in the range 0-2.0.

The spectra of Friedman et al.(2011)
show examples of extra strong (in comparison to E(B$-$V))
5780 DIBs ($\sigma$Sco) or to the opposite case (HD53367).
The scatter observed is caused by different physical properties of
individual clouds. In cases of a
broad E(B$-$V) range, the individual differences become negligible in relation
to E(B$-$V) and thus the correlation starts looking tight. However,
applying such average relation to individual objects is ill-recommended.
These examples confirm the early result of Kre{\l}owski \& Westerlund (1988).
Our relation proves that 5780
DIB is as good a proxy of E(B$-$V) as the Gaia DIB.  Longer sightlines likely intersect many clouds and
thus one observes in every case an ill-defined average; such
averages are similar if the number of intersected clouds is
large (Kre{\l}owski \& Strobel (2012)). This is
why the correlation between DIB intensities and E(B$-$V) is
tighter for distant objects and why DIB intensities
are risky E(B$-$V) proxies in cases of nearby stars.

Plain evidence of this fact can be found in Fig. 6
where spectra of two different stars are shown. It seems to be
interesting that 5780 DIB, which is rather broad,
is relatively weak in the spectrum of HD204827 while the narrow
5797 DIB is especially strong. This resembles the $\zeta$ and
$\sigma$ type clouds, first mentioned by Kre{\l}owski \& Westerlund (1988), and HD219287
resembles the $\sigma$ case. It is to be mentioned that the 8620 DIB
behaves in unison with the 5780 DIB, which is similarly broad.

\begin{figure}
\includegraphics[width=12cm]{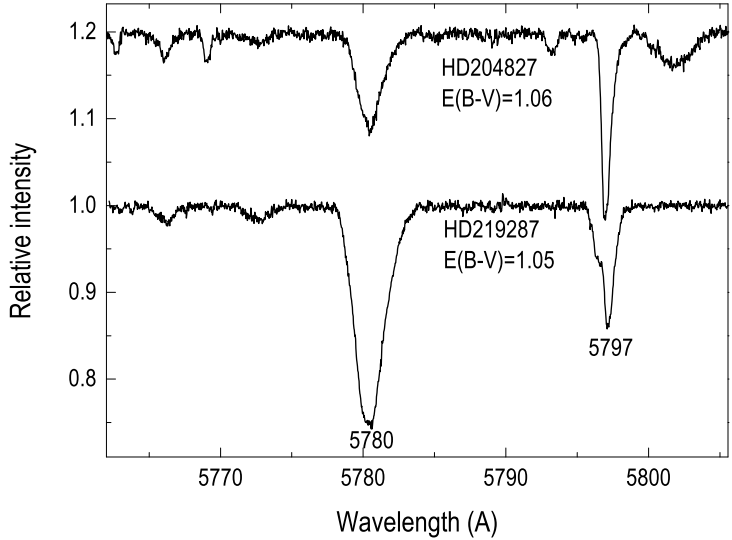}
\caption{The strength ratio of two major DIBs: 5780 and 5797 may be of evidently different strength ratio
in two objects of identical reddening.}
\label{comp5780}
\end{figure}

DIBs are commonly believed to be carried by some interstellar molecules.
However, the molecular species, identified in translucent clouds, where
DIBs are observed, are usually just diatomics; the strongest features
belong to CH and CH$^+$. Are the latter related to E(B$-$V)?
Fig.~7 demonstrates that individual relations between
reddening and strength of molecular features (e.g. CH$^+$) may differ from object to
object.  Apparently, simple interstellar molecules behave in translucent
clouds in the same fashion as diffuse band carriers. However, it does
not mean that the behavior of the latter allows one to predict that of
simple molecules. For instance, the spectrum of HD204827, where 5780 DIB is
very weak, contains extra strong CH lines (like in any $\zeta$ case). We  also
demonstrated recently (Kre{\l}owski et al. 2019) that CH$^+$ lines can be strong
(in the Pleiades cluster) while  all DIBs are barely visible.

\begin{figure}
\includegraphics[width=12cm]{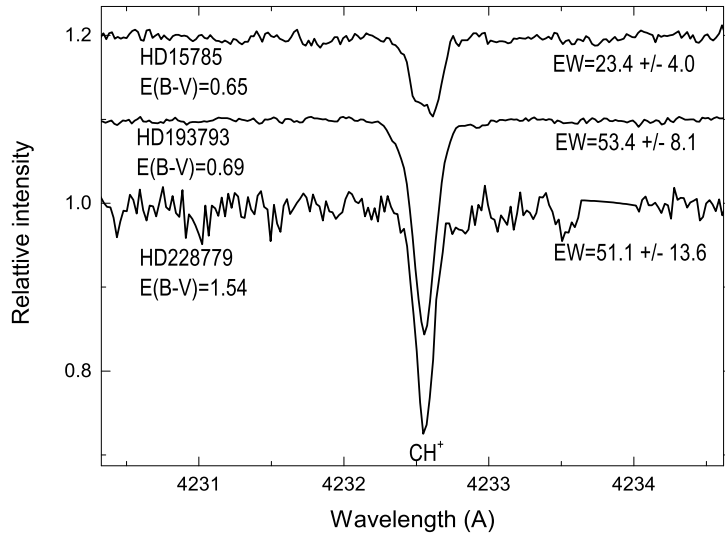}
\caption{Intensities of diatomics may be different with the same E(B$-$V) or identical with different E(B$-$V).}
\label{molint}
\end{figure}


\begin{figure*}
\centering
  \includegraphics[width=14cm]{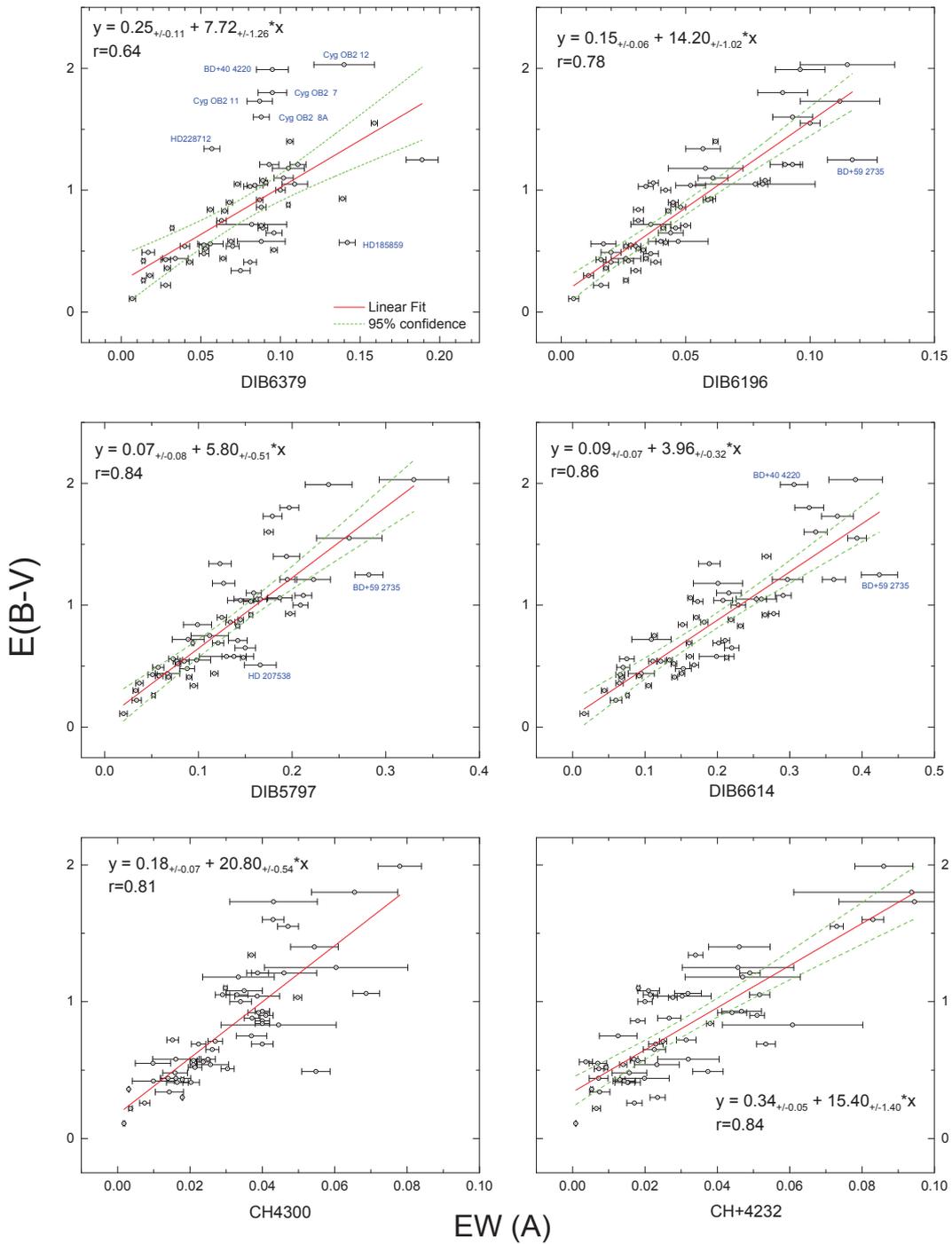}
  \caption{Correlation between E(B$-$V) and equivalent widths of several DIBs
and molecular lines of simple diatomics. The correlations, except for 6379 DIB (see text),
look very similar.}
  \label{dibs_ce}
\end{figure*}

Finally, we have analyzed the correlations of intensities of several
reasonably strong DIBs to E(B$-$V). Figs. 3 and 4, as well as  Fig.~8
demonstrate that in cases
of practically all diffuse bands and features of simple diatomic
molecules, one can build correlations which are of similar
correlation coefficients. The Gaia DIB does not seem to be an
exclusive case. Almost every other interstellar spectral features, only if
measured precisely enough, can be used to estimate E(B$-$V) with a
similar accuracy (and risk). The correlation looks poor in the case of
the 6379 DIB. This is most likely because of the contaminating stellar nitrogen line,
which is difficult to be properly removed even in our high resolution spectra.
In the case of low resolution, the separation is entirely impossible.

\section{Conclusions}

In our research, we used 56 OB stars to measure intensities of several DIBs
in order to check their correlation with color excess E(B$-$V). It is appropriate
to mention here that the relation between reddening and equivalent width of DIBs
has been studied for a long time.
In particular, Munari et al.(2008) proposed to use the 8620 DIB as a tool to estimate
reddening of stars. This diffuse interstellar band at $\lambda$ 8620.7 \AA\
 is the only unambiguous one, which falls into the spectral range
of the Gaia RVS spectrometer.

A major conclusion of our analysis is that DIB carriers are in a way related to
dust grains (spatial correlation?) but none shows a very tight
relation. The 8620 DIB is of importance for observations with Gaia,
but the 5780 DIB gives a similarly accurate E(B$-$V) proxy. It
must be emphasized that an estimate of reddening, based on any
interstellar spectral feature, may be incorrect, despite the rather
tight correlation between any of the interstellar lines/bands and
E(B$-$V) if the range of the latter is broad. Our fits of straight lines were done without fixing the
intercepts. Anyway, the calculated intercepts are nearly zero
indeed. Only in the case of the CH$^+$ line (Fig.~8) we
observe the intercept as big as 0.35. Actually, the line is
observed in practically zero reddening (e.g. in Pleiades).
Apparently, the existing fits depend on the samples selected. On
the other hand, the fact that all interstellar features do
correlate tightly with reddening may suggest that the
carriers of all these lines/bands/extinction are (in average)
reasonably well mixed in all interstellar clouds. Thus, the
observations of identified features of interstellar molecules can
be used to estimate physical properties of clouds and analyze their
influence on diffuse band carriers only in cases of targets being
obscured by single absorption clouds. Since the latter differ from
object to object we observe a very similar scatter of physical
origin in any sample of similar E(B$-$V). Only adding such
``boxes''  we get a correlation. Thus the correlations between
different DIBs and E(B$-$V) look tight in cases of a broad range of
extinction, most likely caused by many clouds along any line-of-sight.
Apparently, such average relations allow estimates of
E(B$-$V)'s from intensities of interstellar spectral features only in cases
of high values of E(B$-$V)'s originated in several clouds each. High extinction,
single cloud objects (like HD204827) are evident outliers from the average relation.

The above considerations lead to the following conclusions:
\begin{itemize}
\item
intensities of all DIBs (not only those of the 8620 DIB) are correlated to a similar degree with E(B$-$V);
\item
the above relation is valid if a large diversity of interstellar clouds is averaged. Individual
clouds may be peculiar but only these peculiarities can be physically interpreted;
\item
the scatter observed in our plots is of physical origin. This is why we observe  tight correlations
only in cases of broad ranges of E(B$-$V)'s resulting from adding up the optical depths of several (many) clouds;
this confirms the conclusions of Bailey et al. (2016) and earlier publications on the subject;
\item
only in the above mentioned cases (high E(B$-$V)'s, many clouds) intensities of DIBs, in particular
of the Gaia DIB at $\lambda$ 8620.7 \AA, may serve as an extinction proxy.
\end{itemize}

\Acknow{
JK acknowledges the financial support of the Polish National Science
Centre (the grant UMO-2017/25/B/ST9/01524 for the
period 2018 \textemdash 2021).
Polish and Ukrainian authors benefited from the funds of the Polish-Ukrainian
PAS/NASU joint research project (2018~-- 2020). GG and JK acknowledge the support of CONICYT program "REDES Internacionales" project
REDES180136.
}

\end{document}